# On the (im)possibility of sustainable artificial intelligence*

## Why it does not make sense to move faster when heading the wrong way

Rainer Rehak

*Weizenbaum Institute for the Networked Society, Berlin, Germany*
*rainer.rehak@weizenbaum-institute.de*

**Abstract: The decline of ecological systems threatens global livelihoods and therefore increases injustice and conflict. In order to shape sustainable societies new digital tools like artificial intelligence (AI) are currently considered a "game-changer" by many within and outside of academia. In order to discuss the theoretical concept as well as the concrete implications of 'sustainable AI' this article draws from insights by critical data and algorithm studies, STS, transformative sustainability science, and more remotely, public interest theory. I argue that while there are indeed many sustainability-related use cases for AI, they are far from being "game-changers" and are likely to have more overall drawbacks than benefits. To substantiate this claim, I differentiate three 'AI materialities' of the AI supply chain: first the literal materiality (e.g. water, cobalt, lithium, energy consumption etc.), second, the informational materiality (e.g. lots of data and centralised control necessary), and third, the social materiality (e.g. exploitative global data worker networks, communities heavily affected by waste and pollution). Effects are especially devastating in the global south while the benefits mainly actualize in the global north. Analysing the claimed benefits further, the project of *sustainable AI* mainly follows a technology-centred efficiency paradigm, although most literature concludes heavy digital rebound effects in the past and also in the future. A second strong claim regarding *sustainable AI* circles around so called apolitical optimisation (e.g. regarding city traffic), however the optimisation criteria (e.g. cars, bikes, emissions, commute time, health) are purely political and have to be collectively negotiated before applying AI optimisation. Hence, *sustainable AI*, in principle, cannot break the glass ceiling of transformation and might even distract from necessary societal change. Although AI is currently primarily used for misinformation, surveillance, and desire creation, I close the article by introducing two constructive concepts for sustainable and responsible AI use, if there is no societal will to refrain from using AI entirely. First, we need to stop applying AI to analyse sustainability-related data, if the related scientific insights available already allow for sufficient action. I call using AI for the sake of creating non-action-related findings *unformation gathering*, which must be stopped. Secondly, we need to apply the *small is beautiful principle*, which means to refrain from using very large AI models and instead turn to tiny models or just to advanced statistics. This approach nicely covers virtually all actual AI use cases, is orders of magnitude less resource hungry and does not promote power centralisation as large models do. This article intends to further the academic critical AI discourse at the nexus between useful AI use cases, techno-utopian salvation narratives, the exploitative and extractivist character of AI and concepts of digital degrowth. It aims to contribute to an informed academic and collective negotiation on how to (not) integrate AI into the sustainability project while avoiding to reproduce the status quo by serving hegemonic interests.**

---

# Introduction

Two of the existential issues of our time are the climate catastrophe and the dramatic loss of biodiversity. They are existential in the way that our human livelihood depends on functioning ecosystems (WBGU, 2019). Therefore, calls for ecological sustainability are in fact calls for protecting human life, democratic societies, and their institutions, such as public interest infrastructures in a globally fair and equitable fashion. The decline of ecological systems directly entails the accelerating magnification of long known political problems like global inequality, poverty, hunger, or war. Increased scarcity of life-sustaining resources, even concerning habitable areas, fuel conflict and injustice (Abel et al., 2019).

In light of the herculean endeavour of shaping sustainable societies (e.g. see the UN Sustainable Development Goals), where the needs of all present human beings as well as future generations are met within the planetary ecological boundaries (Raworth, 2012), promising approaches and solutions are desperately needed. The search especially includes the use of digital tools like artificial intelligence (AI), which is currently considered a "game-changer" and "revolutionary" technology in the fight against climate change by many, even the UN itself (UN-OHCHR, 2024).

While AI is not one single technology but a large bouquet of related digital methods, there are indeed use cases for sustainability-oriented AI systems like circular economy optimisation (Wilson et al., 2022), reduction of resource and energy consumption (Himeur et al., 2021), local CO2 emission reduction (Alli et al., 2023), smart city optimisation (Cugurullo, 2020), sustainable mobility (Vermesan et al., 2021), waste separation and disposal (Wilts et al., 2021), and new ways of information gathering like environmental pollution detection (Pouyanfar et al., 2022) or radically improved approaches to earth observation (Bereta at al., 2018). Calling these and other concrete examples a sustainability game-changer discloses a very reduced understanding of the sustainability project, as well as of AI technology, because it ignores the political and economic context of AI applications and even its own resource demand. I therefore want to argue that centering any AI technology in the sustainability struggle will, if continuing on the current path, very likely do more harm than good and might even be a distraction from the actually relevant societal tasks of a sustainability transformation. I advocate for moving forward in a problem- and goal-oriented manner and not in a technology- or even AI-driven one, as AI might eventually play a role in societal sustainability transitions, although a very small one.

In this short opinion piece I extensively draw from insights by critical data and algorithm studies, STS, transformative sustainability science, critical computer science, and, more remotely, public interest theory. I will first outline some core properties of AI systems, then socially contextualise and critically discuss these regarding their sustainability implications. I will make constructive remarks concerning a sensible use of certain kinds of AI in combating problems like climate change and conclude the piece.

# AI materialities

AI, like all digital tools, has specific common materialities. While the concept of materiality traditionally refers to physical properties, I will use it in a more abstract sense underlining its resource-ness. This includes physical characteristics, but also informational properties and limitations, and even inner and outer social dynamics (Rehak, 2021; Rehak, 2023).

On the physical level it is important to focus on the ecological implications. While there are many possible sustainability-related use cases of AI, the concrete technologies residing under this umbrella term have an astronomically high consumption of material resources like water, cobalt lithium, or energy (e.g., Li et al., 2023; de Vries, 2023), whose production and disposal has major ecological impacts, especially in the global south. Overall, the use of AI skyrockets all indicators of current and future estimated digital ecological impact (Taddeo et al., 2021). Major tech companies like Microsoft and Google just reported rises in resource uses previously unheard of, stopped carbon offsetting, and announced that they would miss their already loose (Hoffmann, 2022) sustainability pledges, precisely because of their large scale AI roll-out (e.g. Marx, 2024; Metz, 2024; Rathi & Bass, 2024). Although there are attempts to minimise AI resource use (Wu et al., 2022; Nenno, 2024), their success is very limited, and efficiency gains will very likely be eaten up by the usual digital rebound effects (Freitag et al., 2021; Lange et al. 2020; Bergman & Foxon, 2022).

Regarding the informational materiality, AI systems can be differentiated between *discriminative* and *generative* systems. The former can detect patterns in input data (e.g. categorisation, grouping), the latter can produce patterns (e.g., text, images) related to input data. However both are based on advanced statistics and provide heuristic (not precise) results. This makes AI systems applicable for specific use cases where clearly defined and precise results are not needed, but AI is in no way a "general purpose" technology (Bender et al., 2021). In addition, AI systems require large amounts of up-to-date valid data to be preconfigured for tasks. This implies a tendency towards centralisation as well as fundamental limits regarding practicability and usefulness (Bender et al., 2021).

And concerning the aspect of social materiality, the global AI supply chain has to be taken into focus. AI inherits the general implications of the digital supply chain, but in a much more intense way. Not only do AI technologies consume vast amounts of materials, lots of them from conflict-torn regions (e.g. conflict minerals, with the mining mainly commissioned by western corporations) and are disposed of in equally problematic e-waste sites heavily affecting the health and freedom of the workers and communities involved. Furthermore, many AI systems require complex and exploitative global networks (Mühlhoff, 2020) of data workers gathering, sorting, categorising, and labelling data in precarious and disempowered work arrangements (Miceli & Posada, 2022), again especially in the global south.

# The purpose of (sustainable) AI

After having briefly outlined some key characteristics of AI, we can now reflect further on its current use and its general potential, beyond the ecology-oriented applications mentioned above. A current hot topic in public and academic AI discourse is how to use "AI technology" for the common good. While the transition from focussing on the values being inscribed into AI systems to focussing more on AI governance itself by applying public interest theory seems very promising for specific public service use cases (Züger & Asghari, 2023), a sustainability perspective has to take an overall *net impact* stance. From this point of view of limited resources, we need to choose wisely where to (not) use AI.

The current overpromises regarding sustainable AI could not be explained by small claims like sorting trash better or optimising the length of machinery use by predictive maintenance alone, but rather by impressive claims like mega city traffic optimisation or global fair resource distribution, meaning: really hard societal problems. But while AI is actually quite useful for sorting trash and predictive maintenance (Wilts et al., 2021; Kamel, 2022), the latter two are, in fact, no technical problems to be calculated, but social problems of collectively agreeing on the very meaning of *optimality* in the given case.

The example of city traffic optimisation exemplifies that there first has to be a decision on what to optimise for, before then using AI based optimisation systems. Traffic could be optimised for cars, bikes or pedestrians, for low emissions, short commute time, good citizen health, or any other parameter or mix of potentially contradicting parameters. AI cannot make this decision, since it has neither agency nor is it neutral (Rehak, 2021; Prietl, 2019). It cannot produce objectively good solutions agreeable by everyone. I leave the task to the reader to apply this thought experiment to the case of global resource distribution.

The main claim of AI, that it could technically produce a result, which is in fact the social precondition necessary to meaningfully apply AI, is clearly just circular reasoning. This misjudgement explains why AI can not be the "game changer" being able to break the glass ceiling of transformation (Hausknost, 2019): AI systems can only implement what we as society have decided so far and are always created under the current, fundamentally unsustainable framework conditions. AI therefore currently, and under current conditions, only helps to carry on with business as usual (cf. Kwet, 2019), and *sustainable AI* only works as change placebo resulting in placebo change; or to put it more concisely: "sustainable AI is the technical solution to the climate crisis from a techno-solutionist vantage point simply reproducing the status quo. The enthusiasm for sustainable AI primarily serves hegemonic interests" (Schütze, 2024, p. 1).

The necessary societal and political negotiations cannot be automated (Eyert & Lopez, 2023), and for the sake of sustainability it is crucial to decenter technological thinking and focus on questions of power and social change (Creutzig et al., 2022).

# A path forward in the public interest

If we do not want to employ sustainability reasons to ban AI altogether, because we still want to harness the impressive results it can have in some areas, two guiding principles should be seriously discussed and taken into account in policy discourse, better sooner than later.

**Stop *unformation* gathering** – Many new sustainability-related AI applications promise to generate new insights, may it be regarding forest health, biodiversity indicators or crop quality. Yet, in most cases of possible protective climate or biodiversity action we as humanity already have sufficient scientific actionable insight. Getting more information is never bad, but the grave implications of AI use might not justify mere theoretical curiosity. Such research could even have a delaying effect, if decision-makers use it to wait for more detailed but practically meaningless results. I call these kinds of results, i.e. the information produced by an obsessive focus on getting more data – without taking action on the sufficient knowledge already available – *unformation*, which must be called out and prevented, especially when used as pretext for inaction.

**Small is beautiful** – The larger an AI system, the more devastating are the sustainability-related effects as described above. Although there is much talk currently regarding large language models (LLMs) like ChatGPT, Bart, and the like, the field of AI itself is already nearly 70 years old and has also, together with adjacent disciplines such as statistics, produced all kinds of rather lightweight approaches (e.g. Hansen et al., 2020). Those small systems are usually really good for one specific use case, like water usage optimisation or visual bike detection. They are not only environmentally lightweight, but also do not promote power centralisation like LLMs necessarily do (Rehak, 2023, p. 30); small AI is beautiful AI. Furthermore, currently all kinds of purposes are being pursued utilising powerful AI systems. However, in many areas it has been proven that with a bit of development effort, similarly usable results can be achieved using conventional and lightweight non-AI data processing methods, such as traditional statistics (e.g. Cerasa et al., 2022). If this is possible, AI must not be used. So let's focus on lightweight systems and call for digital degrowth (Aanestad, 2023; Kwet 2024).

# Collective conclusion

Currently hyped kinds of AI like LLMs are amongst the resource hungriest digital technologies we humans have ever developed and they depend on exceptionally exploitative and destructive AI supply chains, e.g. the broad introduction of AI drives the construction of one hyperscaler data centre after the other (Marx, 2024) while taking renewable energy away from other uses. From a sustainability point of view such technology needs an extremely well argued case for still being deployed today. Merely increasing profit for a few is no such case and given the fact that LLMs are largely used for misinformation, surveillance, and desire creation (Castor & Gerard, 2023), it is hard to see why such systems should exist at all. There are indeed some benefits, but are they really worth it?

No single person can answer that, so this clearly is a topic for collective negotiation, preferably as a public interest issue and preferably without AI stakeholders.

## Funding Note

This work was funded by the Federal Ministry of Education and Research of Germany (BMBF) under grant no. 16DII131 ("Deutsches Internet-Institut").

# About the Author

Rainer Rehak is part of the research group "Digitalization, Sustainability, and Participation" at the Weizenbaum Institute for the Networked Society, he is an associated researcher at the Berlin Social Science Center (WZB) and is currently doing his PhD on systemic IT security and societal data protection at the TU Berlin.

He studied computer science and philosophy in Berlin and Hong Kong and has been working on the implications of the computerization of society for over 15 years. His research fields include data protection, IT security, state hacking, computer science and ethics, fictions of technology, digitization and sustainability, convivial and democratic digital technology, and the implications and limits of automation through AI systems.

He also publishes regularly in non-scientific outlets and is an expert witness for parliaments (e.g., the German Bundestag) and courts (e.g., the German Constitutional Court). Together with other digital policy and environmental organizations, he initiated the "Bits & Bäume" conference for digitization and sustainability.

Profile website: https://www.weizenbaum-institut.de/en/portrait/p/rainer-rehak/

# Licence